\documentclass[apj]{emulateapj}
\usepackage{apjfonts}
\usepackage{graphicx}
\usepackage{amsmath,amssymb}

\def\apj{ApJ}

\def\aap{AAp}

\def\pasp{PASP}

\let\oldAA\AA
\renewcommand{\AA}{\text{\normalfont\oldAA}}

\begin{document}

\title{Galactoseismology: Discovery Of A Cluster of Receding, Variable Halo Stars}
\author {Sukanya Chakrabarti  \altaffilmark{1}, 
Rodolfo Angeloni  \altaffilmark{2}, 
Kenneth Freeman  \altaffilmark{3}, 
Benjamin Sargent  \altaffilmark{1}, 
Joshua D. Simon  \altaffilmark{4}, 
Piotr Konorski \altaffilmark{5}, 
Wolfgang Gieren\altaffilmark{6},
Branimir Sesar  \altaffilmark{7}, 
Andrew Lipnicky  \altaffilmark{1} ,
Leo Blitz  \altaffilmark{8}, 
Gibor Basri  \altaffilmark{8},
Massimo Marengo  \altaffilmark{9}, 
William Vacca \altaffilmark{10},
Puragra Guhathakurta \altaffilmark{11},
Alice Quillen  \altaffilmark{12}, 
Philip Chang  \altaffilmark{13}
}

\altaffiltext{1}
{School of Physics and Astronomy, Rochester Institute of Technology, 84 Lomb Memorial Drive, Rochester, NY 14623; chakrabarti@astro.rit.edu}
\altaffiltext{2}
{Gemini Observatory, c/o AURA, Casilla 603
La Serena, Chile}
\altaffiltext{3}
{Research School of Astronomy \& Astrophysics
ANU RSAA
Mount Stromlo Observatory}
\altaffiltext{4}
{Carnegie Observatories}
\altaffiltext{5}
{Warsaw University Observatory}
\altaffiltext{6}
{Universidad de Concepcion}
\altaffiltext{7}
{MPIA}
\altaffiltext{8}
{UC Berkeley, Astronomy Department}
\altaffiltext{9}
{Iowa State}
\altaffiltext{10}
{NASA Ames}
\altaffiltext{11}
{UCSC, Department of Astronomy}
\altaffiltext{12}
{University of Rochester, Department of Physics and Astronomy}
\altaffiltext{13}
{University of Wisconsin-Milwaukee}

\begin{abstract}

A dynamical characterization of dark matter dominated dwarf galaxies from their observed effects on galactic disks (i.e. Galactoseismology) has remained an elusive goal.  Here, we present preliminary results from spectroscopic observations of three clustered Cepheid candidates identified from $K$-band light curves towards Norma.  The average heliocentric radial velocity of these stars is $\sim$ 156 km/s, which is large and distinct from that of the Galaxy's stellar disk.  These objects at $l \sim 333 ^\circ$ and $b \sim -1 ^\circ$ are therefore halo stars; using the $3.6~\micron$ period-luminosity relation of Type I Cepheids, they are at $\sim$ 73 kpc.  Our ongoing $I$-band photometry indicates variability on the same time scale as the period determined from the $K_{s}$-band light curve.  Distances determined from the $K$-band period-luminosity relation and the 3.6 $\micron$ period-luminosity relation are comparable.   The observed radial velocity of these stars agrees roughly with predictions from dynamical models.  If these stars are indeed members of the predicted dwarf galaxy that perturbed the outer HI disk of the Milky Way, this would mark the first application of Galactoseismology. 

\end{abstract}


\section{Introduction}

The puzzlingly large perturbations in the outer HI disk of our Galaxy (Levine, Blitz \& Heiles 06) may be due to an interaction with a dwarf galaxy (Chakrabarti \& Blitz 09; henceforth CB09).  Underlying the search for this culprit is the hope that one could in principle infer characteristics of the Galactic potential and sub-structure from a dynamical analysis of observed perturbations in the gas or stellar disk of the Milky Way, i.e, Galactoseismology.  

We recently presented evidence of an excess of variables with periods greater than 3 days, and in particular clustered Cepheid candidates at $l \sim 333^\circ$, $b \sim -1^\circ$ (Chakrabarti et al. 2015; henceforth C15) from the analysis of $K_{s}$-band light curves from the VVV survey (Saito et al. 2012).  Using the $K_{s}$-band period-luminosity relation of Type I Cepheids (Matsunaga et al. 2013), and the Cardelli et al. (1998) extinction law, we derived an average distance of $\sim$ 90 kpc from the Galactic Center, which is close to the current predicted distance of the perturbing dwarf galaxy.  

Without both light curves and spectroscopic observations, it is difficult to understand the nature of variable stars and their dynamical relation to our Galaxy. 
Here, we present preliminary results from $H$-band spectroscopic observations of three Cepheid candidates identified from the analysis of $K_{s}$-band light curves from the VVV survey.  The average heliocentric radial velocity of these sources is $\sim$ 156 km/s, which is large and distinct from the Galactic stellar disk (Yanny et al. 2003; Bond et al. 2010).  

We address the recent claims by Pietrukowicz et al. (2015; henceforth P15) that these sources are cool spotted stars, and the broader claims that classification of variables from near-IR light curves is problematic, and that there is no evidence for a dwarf galaxy in this part of the sky.  Given that the period of one of these sources has been confirmed by OGLE, the main false positive are spotted stars that can mimic the $K_{s}$-band light curves of Cepheids.  However, spotted stars would be part of the Galactic stellar disk, while the observed spectra indicate that these objects are moving with a large (and similar) radial velocity, which is much larger than that of typical disk stars.  Moreover, the distance of these sources as derived from the period-luminosity relation in the $K_{s}$-band and at $3.6~\micron$ are comparable.   Our own ongoing I-band photometry indicates variability on roughly the same scale as the period derived from the $K_{s}$-band light curves.  The $I$-band Fourier parameters of one of these sources (P15; Figure 5 in that paper), while not lying in the main part of the distribution of Fourier parameters, shows some overlap with SMC Cepheids, as would be expected for Cepheids in a low-metallicity environment.  Thus, the possibility that this source is a classical Cepheid should not be dismissed.  Finally, the $5-\sigma$ excess of variables in this part of the sky (C15) is difficult to explain without invoking a gravitationally bound structure.

The classification of variable stars using near-IR light curves (Matsunaga et al. 2011; 2013; Angeloni et al. 2014) is indeed a relatively new endeavor.  While the traditional approach has been to classify variable stars in the optical or $I$-band (Soszynski et al. 2011) and obtain a few epochs of near-IR photometry to lower the uncertainty on the distance (Feast et al. 2014), the challenge of classifying sources that are faint in the optical bands motivates the development of templates in the near-infrared (Angeloni et al. 2014; Inno et al. 2015), as well as other innovative approaches of multi-color classification with relatively few epochs (Sesar et al. 2012).   Light curves of Cepheids in the near-IR are nearly sinusoidal (Matsunaga et al. 2012), while light curves in the $I$-band tend to show more features (Soszynski et al. 2011), allowing easier classification.  But even with $I$-band light curves, the classification is not always accurate. Only five of the thirty-two Cepheid candidates identified by Soszynski et al. (2011) were spectroscopically confirmed by Feast et al. (2014) (while the others are not Type I Cepheids or else their identification as classical Cepheids is uncertain).  Given the relative paucity of data in the near-IR, it is reasonable to proceed initially with light curves (with a moderate number of epochs) and subsequently obtain spectroscopic data, especially for sources close to the Galactic plane.  

Studying Galactic structure and sub-structure close to the plane is also a new field.  
Young Cepheid variables (or a metal-rich component) are unexpected in models of the Galactic halo, especially in the outer halo.  However, gas accretion or infall of gas-rich dwarf galaxies may well produce an yet undiscovered population of Type I Cepheids in the outer Galaxy.  Bullock \& Johnston's (2005) cosmologically motivated models of the stellar halo suggest that while most of the satellite accretion occurred early in the Milky Way's past, there may have been some recent accretion events.

\section{Cepheid Candidates}

\begin{figure}
\begin{center}
\includegraphics[scale=0.4]{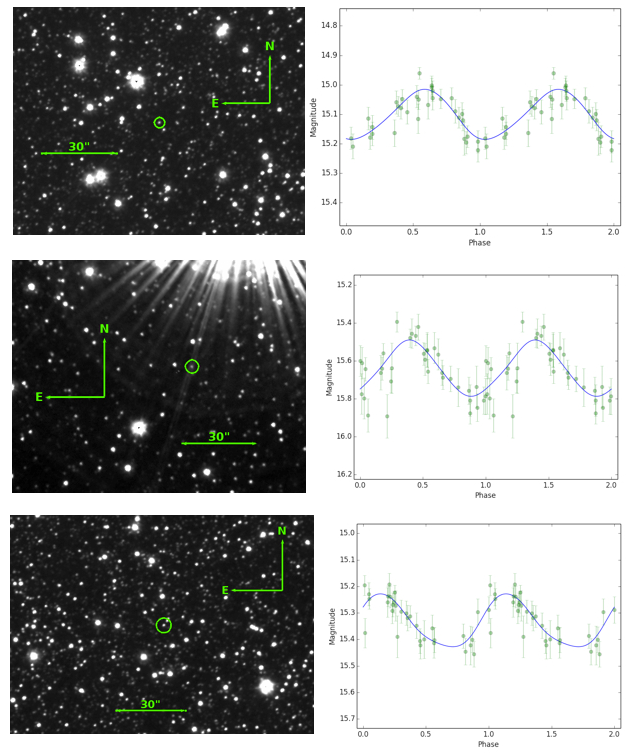}
\caption{H-band images with North up and East to the left, with the horizontal line marking 30" on each plot, and $K$-band light curves of the Cepheid candidates of the (top) P  = 5.69d source, (middle) P = 4.923d source, and (bottom) P = 4.9d source
\label{f:sources}}
\end{center}
\end{figure}

\subsection{Identification of Cepheid Candidates}

As in C15, we derive the periods of the sources presented in Table 1 using the Lomb-Scargle algorithm and retain only sources that have a significance level greater than 90-th percentile (Klein et al. 2014), and periods greater than three days and $K_{s} > 15$ mag. We then perform a parametric bootstrap assuming a Gaussian distribution of errors and at this stage only retain sources where the mean of the period distribution minus the standard deviation exceeds three days. Finally, we compute the Fourier parameters of the sources and visually inspect the light curves. The finding charts and $K_{s}$-band light curves of the three Cepheid candidates that we obtained spectroscopic observations of are shown in Figure 1.  Our selection of Cepheid candidates differs from C15 in one significant way.  In addition to the criteria used by C15,
here we require that at least 90 \% of the $K_{s}$-band VVV photometric
observations are classified as "stellar", i.e. have flag =-1 (Saito et
al. 2012); spurious variables may be identified if this flag
is not required.   This may be the reason why two of the sources
listed in C15 (the P = 3.42d and the P = 4.19d) were misidentified as
variables.   Only one of the sources in this paper was earlier presented in C15 (the P = 5.69d source).

Our own ongoing $I$-band photometry confirms the variablity of the
three sources in this paper on the same temporal scale as the period determined from the $K_{s}$-band light curves.  
The sources we present here have spectroscopic observations, as well as $I$-band photometry (on average $\sim$ 10 epochs),
 $K_{s}$-band light curves, and 3.6 $\micron$ fluxes.  There are other Cepheid candidates in this part of the sky
that we were unable to obtain spectroscopic observations of; the stars discussed in this paper do not represent a complete sample
of Cepheid candidates in this part of the sky.

Table 1 lists the RA and DEC of the sources in J2000, the pulsation period, as well as the distance derived from the $K_{s}$-band period-luminosity relation (M13), the $3.6~\micron$ period-luminosity relation (Scowcroft et al. 2011), the mean $I$-band magnitude (and current number of epochs from our ongoing photometry on the Swope telescope that will be presented in a forthcoming paper), the mean $K_{s}$-band magnitude, and the $3.6~\micron$ flux.  The $3.6~\micron$ photometry for the P = 5.69d and P = 4.9d sources are from the DEEP GLIMPSE survey (Churchwell et al. 2009; Benjamin et al. 2003), and the $3.6~\micron$ photometry for the P = 4.923d source is from Willis et al. (2015).  
Table 1 also lists the extinction at $K_{s}$ and $3.6~\micron$, and the MJD (and phase at time of the F2 observation) as well as our derived velocity, in the heliocentric frame.

\begin{table*}
\centering
        \caption{Data For Individual Cepheid Candidates and Derived Parameters}

          \begin{tabular}{@{}lccccccccccc@{}}
          \hline

RA                    &       DEC                 &    P (day)    &    $D_{\rm K-band}$ (kpc)      &  $D_{3.6~\micron}$ (kpc)    & <I>                           & <$K_{s}$>  & $3.6~\micron$    & $A_{Ks}$   & $A_{3.6~\micron}$  &   MJD (phase)   & $\rm v_{\rm helio} (km/s)$ \\
\hline

245.330825     &   -52.042578       &      5.69       &   65 $\pm 11$                 &     67  $\pm 6$                        &   19.0 $\pm$ 0.2 (25)       &    15.1 $\pm 0.02$       &   14.55 $\pm 0.07$    & 0.76 &    0.48   &    57293.8 (0.1)   & 162 $\pm 31$   \\

244.968946     &    -50.176306      &      4.923     &  ---                                 &     73 $\pm 20$                       &   $19.0 \pm 0.2$ (7)      &    15.6  $\pm 0.039$      & 15.53  $\pm 0.11$                 & 1.71 &    1.08   &   57278 (0.6) &   151 $\pm 39$ \\

243.140542     &    -54.022122      &      4.9         & 78 $\pm 7$                      &     78 $\pm 6$                       &  $18.9 \pm 0.2$ (8)      &    15.3   $\pm 0.02$     &   14.89  $\pm 0.08$          & 0.45  &    0.28   &  57295 (0.85)   &   154 $\pm 113$  \\

\hline

\end{tabular} 

\small {<I> and <$K_{s}$> refer to the mean I-band and $K_{s}$ band magnitudes respectively from our ongoing observations on the Swope telescope and the VVV survey (Saito et al. 2012), respectively.  3.6 $\micron$ fluxes are from the DEEP GLIMPSE survey (Churchwell et al. 2009; Benjamin et al. 2003) and from Willis et al. 2015.  MJD is the modified Julian date at the time of the F2 spectroscopic observations and the pulsation phase is listed in parenthesis.  Heliocentric velocities are derived from $H$-band F2 spectra presented in this paper.}

\end{table*} 

\subsection{Distances \& Fourier Parameters}

\begin{figure}
\begin{center}
\includegraphics[scale=0.3]{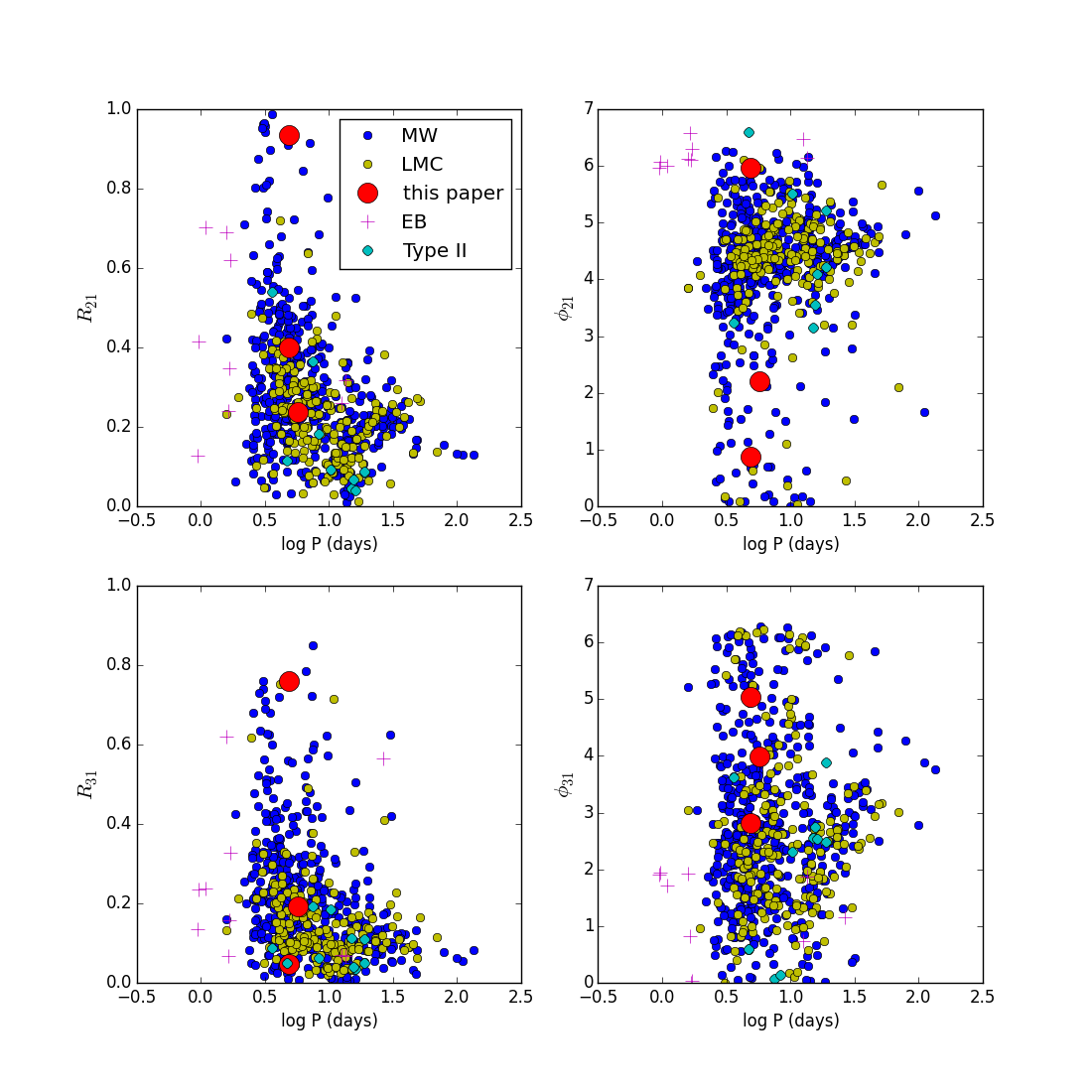}
\caption{Fourier parameters of the Cepheid candidates from $K_{s}$-band light curves, overplotted with Fourier parameters of Type I Cepheids in the Milky Way and LMC from Bhardwaj et al. (2015), and Type I, Type II, and eclipsing binaries (EB) from Matsunaga et al. (2012)
\label{f:fourier}}
\end{center}
\end{figure}

We adopt the period-luminosity relations of classical Cepheids in the LMC (Matsunaga et al. 2011), with a LMC distance modulus of 18.5 mag (Freedman et al. 2001) and interstellar extinction value of $A_{K_{s}}$ = 0.02 mag for the LMC direction (Caldwell et al. 1985). This gives the distance modulus $\mu$ for a Cepheid with pulsation period P :

\begin{equation}
\mu= K_{s} - A_{K_{s}} + 3.284~\rm log(P) + 2.383 \; ,
\end{equation}
where $A_{K_{s}}$ is the extinction in the $K_{s}$ band, which is given in Table 1.
We adopt the "no cut" $3.6~\micron$ period-luminosity relation from Scowcroft et al. (2011) to derive distances from the $3.6~\micron$ photometry, which gives the distance modulus:
\begin{equation}
\mu = m_{3.6} - A_{3.6} + 3.15(\rm log P - 1) + 5.83  \; ,
\end{equation}
where $m_{3.6}$ is the observed $3.6~\micron$ flux and $A_{3.6}~\micron$ is the extinction at $3.6~\micron$.
We use the 90th percentile RGB extinction map from Nidever et al. (2012) to determine $A_{K_{s}}$ whenever available (for the P = 4.923d and P = 4.9d sources), otherwise we use the Bonifacio et al. (2000) correction 
to Schlegel et al. (1989) in conjunction with Cardelli et al. (1989) to get $A_{K_{s}}$ (for the P = 5.69d source).  We then use the Flaherty et al. (2007) extinction law towards
Serpens to get $A_{3.6~\micron}$ from $A_{K,s}$ for all the sources.  
 The sources of uncertainty in the distance are due to the photometric error, the uncertainty in the extinction (taken to be the standard deviation of the extinction values from Schlegel et al. 1998, Nidever et al. 2012 and Schlafly et al. 2011).  For single-epoch measurements of the $3.6~\micron$ flux, the (small) variability (Scowcroft et al. 2011) relative to the observed flux also contributes an uncertainty that is comparable to the photometric error.   We calculate the uncertainty on the distance by adding in quadrature the uncertainties due to these terms.  The distance error derived from the $K_{s}$-band for the P = 4.923d source is formally comparable to the distance itself.  This source is the one that is closest to the plane, and even in the $K_{s}$-band, the extinction values vary significantly.  For this source, we only cite a distance from the $3.6~\micron$ relation.  Given that the $3.6~\micron$ flux has the least extinction and thereby provides the best determination of the distance, this correspondence is reassuring.  If instead these sources were Type II Cepheids, using the period-luminosity relation for Type II Cepheids (Matsunaga et al. 2013), they would lie at $\sim$ 30 kpc from the Galactic center.


Figure \ref{f:fourier} depicts the Fourier parameters (as defined in Equation 3 of C15) of the Cepheid candidates over-plotted with the Fourier parameters of classical Cepheids in the Milky Way and the LMC and SMC from a compilation by Bhardwaj et al. (2015), as well as Type I and Type II Cepheids, and eclipsing binaries from Matsunaga et al. (2013).  The Fourier parameters of Cepheids in the near-IR cover a wider range than the Fourier parameters in the $I$-band, and there is overall agreeement of the Fourier parameters of the Cepheid candidates with known sources in the near-IR, which have nearly sinusoidal shapes.  Our ongoing $I$-band observations indicate that the P = 5.69d and P=4.9d sources show variability on the same scale as the period determined from $K_{s}$-band light curves, but at the moment we cannot compute Fourier parameters in the $I$-band due to the small number of epochs and large photometric errors.  We also cannot yet confirm the level of variability for the P = 4.923d source in the $I$-band.

\section{Follow-Up Spectroscopic Observations}

\begin{figure}
\begin{center}
\includegraphics[scale=0.485]{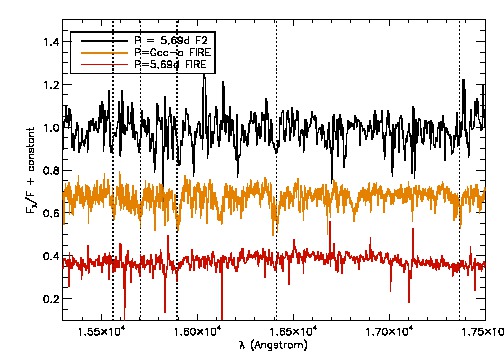}
\includegraphics[scale=0.485]{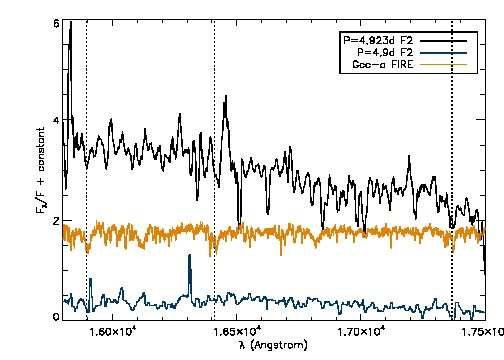}
\includegraphics[scale=0.485]{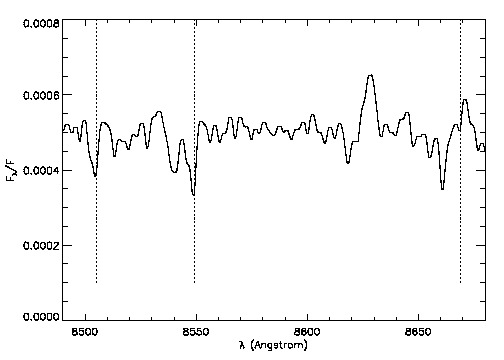}
\caption{(a) H-band spectra of the P = 5.69d source on F2 and on FIRE, and the Gcc-a (the Cepheid from M15) spectrum on FIRE, (b) H-band spectra of the P = 4.923d and the P = 4.9d source on F2 and the Gcc-a spectrum on FIRE, (c) WiFeS spectrum of the P = 4.9d source close to the wavelength region of the expected Calcium triplet lines.  The spectra are shown here median-filtered.
\label{f:spectra}}
\end{center}
\end{figure}

We performed near-infrared spectroscopic follow-up of the three Cepheid candidates with Flamingos-2 (F2) (Eikenberry et al. 2004) on Gemini-South under the DDT Program GS-2015B-DD2. F2 was used in the long-slit mode with the R3K grism and the H filter, thus covering a wavelength range from 1.5 to 1.75 $\micron$. We used the 2-pix (0.36 arcsec) slit for an average spectral resolution R$\sim$2900. The effective on-source total exposure time was 4800 sec per target.  The seeing was better than 0.6 arcsec for the P = 4.9d and P = 4.923d source, and better than 0.85 arcsec for the P = 5.69d source.  We also observed two bright B9V to A0V stars, selected close in time and airmass to our targets to deal with the telluric lines.  The data reduction has been performed with the GEMINI IRAF Package (v1.13) following the standard procedures, and the telluric correction was done following Vacca et al. (2003).  The P = 5.69d source was also observed on the FIRE spectrograph on Baade/Magellan (Simcoe et al. 2013) with an effective on-source exposure time of 1200 sec, and a telluric standard was also observed.  In addition, we obtained a spectrum of the Gcc-a Cepheid on FIRE that was earlier observed by Matsunaga et al. (2015).  The FIRE GUI was used to perform the data reduction. The P = 4.9d source was also observed on WiFeS with an effective exposure time of 4000 sec.  The seeing was about 1.5 arcsec 
and the data were taken in the red channel using the I7000 grating which
gives a nominal resolving power of 7000. 

The $H$-band spectra of Cepheids (and of other F stars) show stronger Brackett series lines and weaker metal lines (Meyer et al. 1998; Rayner et al. 2009) relative to cooler stars.  In particular, the spectra of M dwarfs are distinct from the spectra of Cepheids, while the spectra of cool giants have some correspondence with the spectra of hot supergiant stars.  The $S/N$ of the spectra obtained on Gemini's F2 of these three stars are $\sim 20, 10, 3$ respectively.  Thus, while the spectra are not of sufficient $S/N$ to identify the spectral type of the star, they should be sufficient to measure the radial velocity.   For the Gcc-a Cepheid from Matsunaga et al. (2015), we obtained a similar radial velocity as their listed value from cross-correlation.  Figure \ref{f:spectra} (a) shows the $H$-band spectra of the P = 5.69d source compared with the Gcc-a Cepheid, with the higher $S/N$ lines marked; in both the F2 and FIRE spectra the 14-4 (1.588$~\micron$)  and 12-4 (1.64$~\micron$) transitions are visible.  The higher $S/N$ F2 spectrum also shows the 16-4 (1.556$~\micron$) and 15-4 (1.57$~\micron$) transitions.  Figure \ref{f:spectra} (b) shows the spectra of the P = 4.923d and P = 4.9d source.  The clearest lines for the P = 4.923d source are the 14-4 and 10-4 (1.734$~\micron$) transitions, while for the P = 4.9d source the 10-4 transition appears to be visible.  Due to the relatively low $S/N$ of these spectra and strong telluric lines that overlap with the Brackett lines (especially in the longer half of the $H$-band), not all the Brackett lines can be identified.  To determine the effect of the $S/N$ on the radial velocity, we added noise and degraded the spectrum of the Gcc-a Cepheid and found that even at $S/N \sim$ few, the inferred velocity remained unchanged.  

\subsection{Radial Velocities}

We use a $\chi^{2}$ fitting routine in conjunction with cross-correlation to determine the best-fit velocity and the uncertainty in the velocity.  Our procedure for determining radial velocities is similar to Matsunaga et al. (2015; henceforth M15) who obtained $H$-band spectroscopic observations of classical Cepheids towards the Galactic center.   They have addressed two of the significant challenges in interpreting $H$-band spectra, namely the presence of telluric features in this part of the spectrum, and the relative lack of near-IR radial velocity templates.  To address the first issue, they incorporate the telluric features in the synthetic template spectra developed by Kurucz (1993) (by multiplying the telluric spectrum with the template).  The model spectrum is shifted by a trial redshift and compared with the target spectrum.  Finally, the target spectrum is assigned the redshift that minimizes the $\chi^{2}$ between the shifted model spectrum and the target spectrum.   

Cepheid pulsations can affect the radial velocities by tens of km/s (Pejcha \& Kochanek 2012), and we correct for the pulsation using radial velocity templates at the phase determined from the $K$-band light curves.  We adopt the radial velocity templates used by M15, namely the infrared photometry and radial velocities compiled by Groenewegen (2013).   For Delta Cep, which has a period similar to the stars we consider here, the typical pulsation amplitude is $\sim$ 20 km/s (Barnes et al. 1987).

For the synthetic spectra, we use the following parameters: $\rm T_{\rm eff} = \rm 5000~ K, \rm log~g = 1.0$ and $Z=Z_{\odot}$.   To account for the telluric lines, the synthetic templates incorporate the telluric lines following earlier work in M15.   We also compare to observed templates from the Infrared Template Factory (IRTF; Rayner et al. 2009), as well as our own observation of Gcc-a.   We consider the errors in the spectrum and use a $\chi^{2}$ fitting routine in conjunction with cross-correlation to derive the uncertainties on the radial velocity (following earlier work in Bhalerao et al. 2012 and the implementation in the IDL routine GETVEL.pro).

The heliocentric velocity that we determine for the three stars is $162 \pm 31 \rm ~km/s$, $151 \pm 39 \rm ~km/s$, and $154 \pm ~113 \rm ~km/s$, after correcting for phase using known templates for the radial velocity of Cepheids (Groenewegen 2013; Barnes et al. 1987), as well as comparison with our own observation of Gcc-a on FIRE.  An alternate approach is to mask regions of high sky contamination and perform a cross-correlation, which yields values that are roughly comparable.  The F2 spectrum of the P = 4.9d source is quite low S/N.  The WiFeS observation of the P = 4.9d source produced $\sim$ 10 counts per pixel, and there appear to be Calcium triplet lines that would correspond to the lines at 8498 $\mathrm{\AA}$ and 8542 $\mathrm{\AA}$ (redshifted by $\sim$ 200 km/s) as shown in Figure \ref{f:spectra} (c); however, the redshifted line that would correspond to the 8662 $\mathrm{\AA}$ line is not visible (the latter is in a forest of emission lines).  Due to the low S/N in both the F2 and WiFeS observations for the P = 4.9d source, our radial velocity measurement for this source is particularly uncertain.  

\section{Discussion \& Conclusion}

In this initial study, we were only able to procure spectroscopic observations of three sources.  The excess of $P > 3$d variables in this part of the sky found by C15 is still present when we use the criteria stated in this paper, and future work should attempt to obtain radial velocities of other Cepheid candidates.  In a forthcoming paper, we discuss the excess in the $K_{s}$, $J-K_{s}$ CMD when we extinction correct and subtract out a background field, which is consistent with the location of red-clump stars at $\sim$ 80 kpc.  

The average heliocentric radial velocity of the sources in this paper is $\sim$ 156 km/s.  This radial velocity is large and distinct from the rotation of the stellar disk of our Galaxy, for which |$v_{r}| \sim $ few km/s (Bond et al. 2010).   In this part of the sky, the velocity from the HI map is negative (Kalberla et al. 2005), and from models of the stellar disk (e.g. Besancon et al. 2003) is $\sim -40~\rm km/s$.  Thus, we conclude that, while the identification of these stars as Type I Cepheids may not be certain, these are halo stars, and not part of the Galactic disk.    Interestingly, the average radial velocity is roughly consistent with the predictions from earlier dynamical models (CB09).  This interaction also produces structures in the stellar disk reminiscent of the Monoceros Ring (Newberg et al. 2002), raising the question if the structures in the gas disk and the Monoceros Ring are connected. 


\bigskip
\bigskip

SC is indebted to N. Matsunaga for many helpful discussions on interpreting the spectra.  We thank M. Richmond, N. Morrell, and B. Madore for help with the $I$-band photometry, and D. Newman and R. Simcoe for help with the FIRE pipeline.  We also thank J. Kartaltelpe, V. Scowcroft, and L. Inno for helpful discussions.  
SC is supported by NSF grant 1517488, BS by NASA-ADAP grant NNX15AF15G.

\end{document}